\documentclass{aa}  
\makeatletter
\renewcommand*\aa@pageof{, page \thepage{} of \pageref*{LastPage}}
\makeatother

\newcommand{\Tasc}{T_{\rm asc}}
\usepackage{graphicx}
\usepackage{xcolor}
\usepackage{txfonts}
\usepackage{hyperref}
\def\lum {\mbox{erg\,s$^{-1}$}}

\begin{document}

   \title{UV and X-ray pulse amplitude variability in the transitional millisecond pulsar  PSR\,J1023+0038}


\author{A.~Miraval Zanon\inst{1}, F.~Ambrosino\inst{1,}\inst{2,}\inst{3}, F.~Coti Zelati\inst{4,}\inst{5,}\inst{6}, S.~Campana\inst{6}, A.~Papitto\inst{1}, G.~Illiano\inst{1,}\inst{7},  G.~L.~Israel\inst{1}, L.~Stella\inst{1}, P.~D'Avanzo\inst{6}, M.~C.~Baglio\inst{8,}\inst{6}}

\institute{INAF, Osservatorio Astronomico di Roma, Via Frascati 33, 00078 Monteporzio Catone, Roma, Italy\\
  e-mail: \texttt{arianna.miraval@inaf.it}
  \and INAF, Istituto di Astrofisica e Planetologia Spaziali, Via Fosso del Cavaliere 100, 00133 Roma, Italy
  \and Sapienza Universit\`a di Roma, Piazzale Aldo Moro 5, 00185 Roma, Italy
  \and Institute of Space Sciences (ICE, CSIC), Campus UAB, Carrer de Can Magrans s/n, 08193 Barcelona, Spain
  \and Institut d'Estudis Espacials de Catalunya (IEEC), Carrer Gran Capit\`a 2--4, 08034 Barcelona, Spain
  \and INAF, Osservatorio Astronomico di Brera, Via E. Bianchi 46, 23807 Merate, LC, Italy
  \and Tor Vergata Universit\`a di Roma, Via della Ricerca Scientifica 1, 00133 Rome, Italy
  \and Center for Astro, Particle and Planetary Physics, New York University Abu Dhabi, PO Box 129188, Abu Dhabi, UAE
}

   \date{}
\authorrunning{A. Miraval Zanon et al.}
             
\bibliographystyle{aa}
 
  \abstract
  {The transitional millisecond pulsar PSR\,J1023+0038 is the first millisecond pulsar discovered to emit UV and optical pulses. Here we present the results of the UV and X-ray phase-resolved timing analysis of observations performed with the Hubble Space Telescope, \textit{XMM-Newton}, and NuSTAR satellites between 2014 and 2021. Ultraviolet pulsations are detected in the high luminosity mode and disappear during low and flaring modes, similar to what is observed in the X-ray band. In the high mode, we find variability in both the UV and X-ray pulse amplitudes. The root mean square pulsed amplitude 
  in the UV band ranges from $\sim$2.1\% down to $\sim$0.7\%, while it oscillates in the interval $5.5-12\%$ in the X-ray band. This variability is not correlated with the orbital phase, like what has been observed in the optical band. Notwithstanding
the rather low statistics, we have marginal evidence that variations in the pulse amplitude do not occur simultaneously in the UV and X-ray bands. When the UV pulsed amplitude decreases below the detection threshold, no significant variation in the X-ray pulsed amplitude is observed. These oscillations in the pulse amplitude could be caused by small random variations in the mass accretion rate leading to a variation in the size of
the intra-binary shock region. Finally, we find that the pulsed flux spectral distribution from the X-ray to the UV band is well fitted using a power-law relation of the form $\nu F_{\nu}^{pulsed} \sim \nu^{0.4}$. This supports the hypothesis of a common physical mechanism underlying the X-ray, UV, and optical pulsed emissions in PSR\,J1023+0038.}

   \keywords{pulsars: individual: PSR J1023+0038 -- X-rays: binaries -- stars: neutron}

   \maketitle
%

\section{Introduction}

Transitional millisecond pulsars (tMSPs) are very fast rotating, weakly magnetised (10$^8$--10$^9$\,G) neutron stars (NSs) hosted in binary systems. They attain fast rotations of up to several hundred times per second during a 10$^8$--10$^9$\,year-long phase dominated by disc accretion of matter from a low-mass ($<$ 1\,M$_{\odot}$) companion star \citep{Alpar, Radhakrishnan}. When the mass transfer ceases, these sources shine as rotation-powered radio and/or gamma-ray pulsars, in which the pulsed emission is generated by magnetospheric particle acceleration. The tMSPs reach this regime when the pulsar wind sweeps away the gas spilling out from the companion star through a Roche-lobe overflow. When the mass accretion increases and the NS magnetic field is strong enough to channel the in-flowing matter towards its magnetic poles, the NS can be detected as an accretion-powered millisecond pulsar (MSP). Three confirmed tMSPs are currently known \citep{Papitto_deMartino}: IGR\,J18245–2452 \citep{Papitto2013}, PSR\,J1023+0038 \citep{Archibald, Archibald2013}, and XSS\,J12270--4859 \citep{Bassa2014, deMartino2014}. 

During the accretion-powered state, PSR\,J1023+0038 (J1023 hereafter) and XSS\,J12270--4859 show an X-ray luminosity ($L_X \sim 7 \times 10^{33}$\,erg\,s$^{-1}$, 0.3--79\,keV) that is lower than the value typically observed for accreting MSPs in the outburst phase ($L_X \sim 10^{36}-10^{37}$\,erg\,s$^{-1}$; \citealt{Patruno2021, DiSalvo2019}). During this particular state of tMSPs, which we call the sub-luminous disc state, it is still unclear whether the pulsed emission is powered by the accretion or rotation of the NS magnetic dipole \citep{Papitto2019, Veledina, Campana2019, Miraval2020}. The tMSPs alternate between periods in which the system is brighter in the optical, X-ray, and $\gamma$-ray bands to periods of quiescence during which the radio pulsar is active. The alternation of these states can occur on very short timescales of weeks to months in response to variations in the mass accretion rate. This phenomenology is evidence of a direct link between NS low-mass X-ray binaries and binary millisecond radio pulsars.

J1023 is the only tMSP currently in the sub-luminous disc state and also the best studied source due to its proximity ($d=1368^{+42}_{-39}$\,pc; \citealt{Deller2012}) and brightness ($g\sim16.7$\,mag; \citealt{CotiZelati2014}). During the sub-luminous disc state, J1023 oscillates very rapidly between three different luminosity modes, dubbed low, high, and flaring (\citealt{Bogdanov2015, Campana_DiSalvo, Papitto_deMartino}). The transition between the high, low, and flaring modes is clearly visible in the X-ray and UV bands and occurs simultaneously \citep{CotiZelati2018, Papitto2019, Jaodand2021}. In the optical and near-infrared bands, the low-high mode transition has never been detected, but the flaring activity is clearly visible and could be related to the emission of collimated jets or outflows (\citealt{Shahbaz2018, Papitto2019, Baglio2019}). The low-high X-ray and UV mode transition is anti-correlated with the radio emission \citep{Bogdanov2018}. The low X-ray modes are associated with an increase in the radio activity, which is probably due to episodes of low-level accretion rates and rapid ejections of plasma by the active rotation-powered pulsar \citep{Bogdanov2018}.

During the high luminosity mode, J1023 shows X-ray, UV, and optical pulsations at the NS spin period of $\sim$ 1.69\,ms with a root mean square (rms) pulsed fraction of $\sim8\%$, $\sim0.8\%$, and $\sim1\%$, respectively \citep{Ambrosino, Papitto2019, Jaodand2021}. The low mode is characterised by the absence of X-ray, UV, and optical pulsations. The rms upper limits in the X-ray and optical bands are $\lesssim$ 2.4\% and $\lesssim$ 0.034\% \citep{Archibald2015, Jaodand, Papitto2019}, respectively. During the flaring activity, a decrease in the optical pulse amplitude is observed (0.16(2)\%), and X-ray pulsations are not detected (rms pulsed fraction $<$ 1.3\%; \citealt{Papitto2019}).

In this paper we investigate the UV and X-ray pulse amplitude variability in the tMSP J1023, employing simultaneous or quasi-simultaneous X-ray observations. Using Hubble Space Telescope (HST), \textit{XMM-Newton}, and NuSTAR observations carried out during the sub-luminous disc state, we perform a UV and X-ray phase-resolved timing analysis. This multi-wavelength study probes the properties of UV and X-ray pulses in J1023 along the orbit and in the different intensity modes. We investigate the presence of a relation between X-ray and UV pulsations, confirming their common physical origin.
Section\,\ref{Obs} is dedicated to the description of the dataset and data reduction. In Sect.\,\ref{Timing} we present an X-ray and UV timing analysis as well as the upper limits on the rms amplitude of UV pulses. Section\,\ref{SED} is dedicated to the spectral energy distribution (SED) of the pulsed emissions in the UV and X-ray bands. In Sect.\,\ref{Discussion} we discuss the results and the implication of the UV and X-ray pulse amplitude variability.

\section{Observations and data analysis}
\label{Obs}
In this section we describe the data analysis of the different datasets acquired (quasi-)simultaneously in the X-ray and UV bands over a time span of 8 years. Table\,\ref{table:1} summarises the observations presented and analysed in this paper.

\subsection{UV observations}
\begin{figure*}
   \centering
   \resizebox{\hsize}{!}{\includegraphics{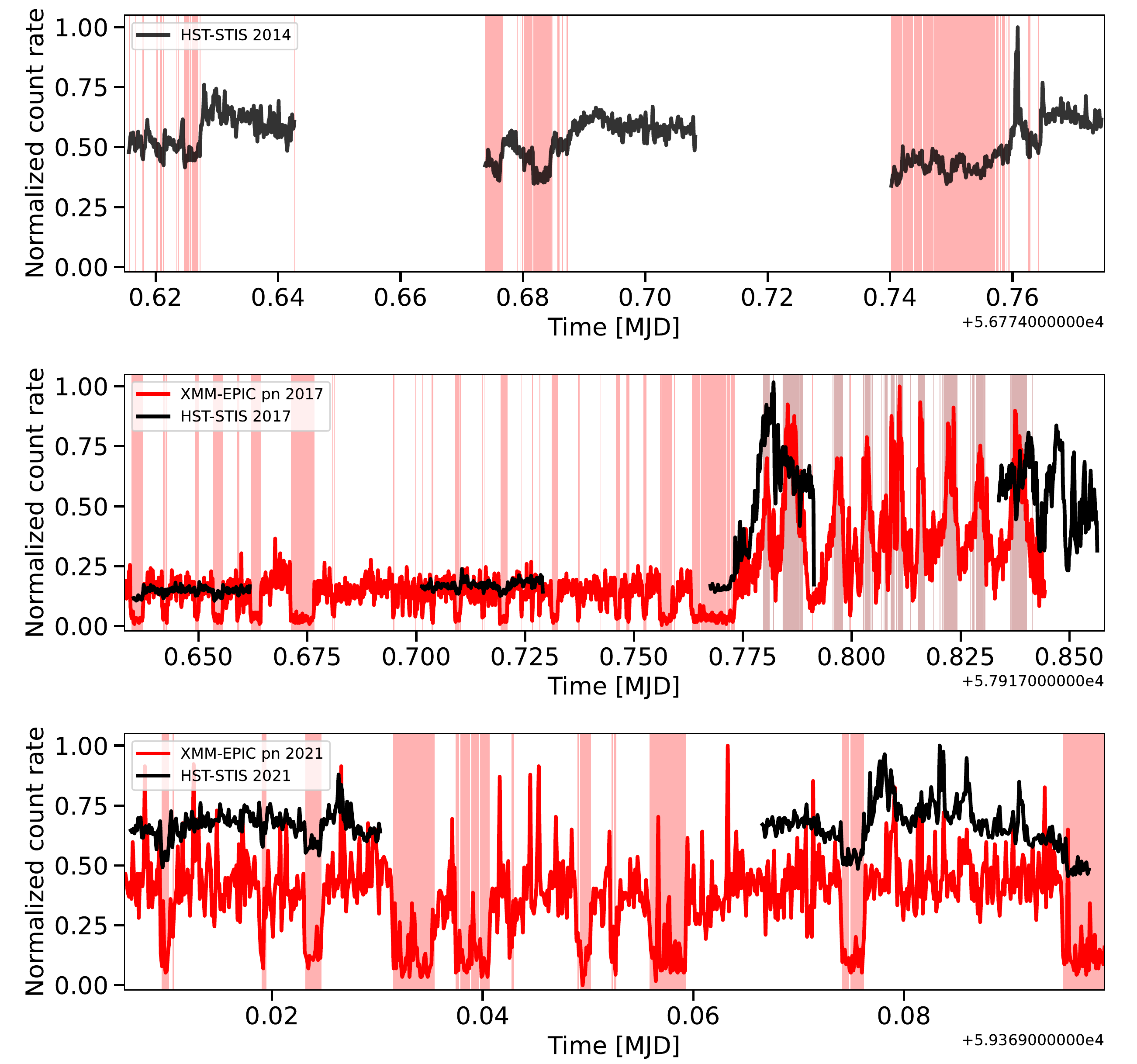}}
   \centering
      \caption{STIS and EPIC-pn light curves observed in 2014 (top panel), 2017 (central panel), and 2021 (bottom panel). Ultraviolet and X-ray light curves are binned every 10\,s. The count rate is
normalised for each instrument at the maximum count rate. The vertical regions indicate the high (white), low (light red), and flaring (dark red) X-ray modes. }
         \label{LC}
   \end{figure*}

The Space Telescope Imaging Spectrograph (STIS) on board HST observed J1023 three times between 2014 and 2021. Details on the nine spectroscopic observations collected during the three HST visits are reported in Table\,\ref{table:1}. These observations were performed in TIME-TAG mode by means of the NUV-MAMA detector with 125\,$\mu$s time resolution; they were acquired with the G230L grating equipped with a 52 $\times$ 0.2\,arcsec slit with a spectral resolution of $\sim$ 500 over the nominal range (first order). We employed the \texttt{stis$\_$photons} package\footnote{\url{https://github.com/Alymantara/stis_photons}} to correct the position of slit channels and assigned the wavelengths to each time of arrival (ToA). We selected ToAs belonging to channels 993--1005 of the slit and in the 165--310\,nm wavelength interval to isolate the source signal, minimise the background contribution, and avoid noisy contribution due to the poor response of the G230L grating at the edge wavelengths. We then referred the ToAs to the Solar System barycentre by using the \texttt{ODELAYTIME} task (subroutine available in the IRAF/STDAS software package), running the JPL DE200 ephemeris with the position of the pulsar reported by \cite{Deller2012}. 
In Fig.\,\ref{LC} we report the normalised UV light curves observed during the three HST visits, binned in time bins of 10\,s. The count rate is normalised for each instrument and year of observation at the maximum count rate, and low, high, and flaring mode intervals are identified. For the observations performed in 2017 and 2021, the selection of the mode intervals was made thanks to the simultaneous \textit{XMM-Newton} X-ray observations (see Sect.\,\ref{X-ray}). For the 2014 dataset we built the distribution of UV count rates by defining the interval 0--76.7\,counts\,s$^{-1}$ as low mode and the interval with count rates $>$ 88.1\,counts\,s$^{-1}$ as high mode because of the absence of strictly simultaneous X-ray observations. The mode selection in the 2014 UV observations is less accurate than that for the observations performed in 2017 and 2021 (see Fig.\,\ref{LC}) because the transitions between the high and low modes (and vice versa) in the UV band are not as evident as in the X-rays.  

\subsection{X-ray observations}
\label{X-ray}

\textit{XMM-Newton} observed J1023 on 2021 June 3--4 (Obs ID. 0864010101) using the EPIC-pn in fast timing mode and both MOSs in full frame mode (see Table\,\ref{table:1}). Data were processed and analysed using the Science Analysis System (SAS; v. 19.1). The net effective exposures after screening for episodes of high background flaring are $\sim$58.7\,ks for EPIC-pn, $\sim$58.1\,ks for EPIC-MOS1, and $\sim$53.5\,ks for EPIC-MOS2. Photon ToAs were barycentred to the Solar System reference frame using the JPL DE200 ephemeris to be consistent with HST observations. Similar procedures as described in \citet{CotiZelati2018} were adopted for the subsequent analysis: source photons were extracted from a 10-pixel-wide strip centred on the brightest pixel column for the pn, and from a circle with a radius of 36 arcsec centred on the source position for each MOS. The background level was estimated from a 3-pixel-wide strip far from the source position for the pn, and from a circle with a radius of 36\,arcsec on the same charge-coupled device (CCD) as the source for each MOS. The time intervals associated with each X-ray mode were selected by extracting the combined background-subtracted light curves from the three EPIC detectors (during the time span of simultaneous coverage) and by applying the same count rate thresholds used by \citet{CotiZelati2018}. The light curves of each EPIC
camera were binned at 10 s. The high mode 
 is defined as when the interval was 4--11 counts s$^{-1}$, the low mode when the
count rate dropped below 2.1 counts s$^{-1}$, and the flaring mode when the count rate exceeded 15 counts s$^{-1}$ (see also \citealt{Archibald2015, Bogdanov2015}). This allowed us to single out the EPIC-pn source event file associated with the high mode for the timing analysis. An analogous procedure was also applied to the datasets acquired in June 2014 and June 2017 (see Table\,\ref{table:1}). For the 2021 datasets, background-subtracted spectra for the time-averaged and high-mode emissions, and associated ancillary and response files, were extracted for both EPIC-MOSs using standard prescriptions (EPIC-pn data were not considered for the spectral modelling due to the known calibration issues at low energies in the fast timing mode).

NuSTAR observed J1023 for an elapsed time of $\sim$48.6\,ks and a net exposure of $\sim$22\,ks (Obs ID. 30601005002), almost fully overlapping with the \textit{XMM-Newton} observation. Data were processed and analysed using NuSTARDAS v1.9.2. The photon ToAs collected by the two focal plane modules (FPMA and FPMB) were barycentred in the same way as for the \textit{XMM-Newton} data. A circle of radius 50\,arcsec (centred on the source) and a circle of radius 100\,arcsec (far from the source and on the same CCD) were used to collect source and background photons, respectively. J1023 is detected up to energies of $\sim$50\,keV in both modules. The time intervals associated with the high X-ray mode derived from the analysis of the \textit{XMM-Newton} data were used to extract the high-mode event files for both modules. The FPMA and FPMB source event files associated with the high mode were then combined to increase the counting statistics for the timing analysis. Background-subtracted spectra were extracted for the whole dataset as well as for the high mode. 

 The \textit{XMM-Newton}/EPIC-MOS and NuSTAR/FPM spectra were modelled together over the 0.3--50\,keV energy range using an absorbed power-law model within \textsc{xspec}\footnote{A re-normalisation factor was also included to account for intercalibration uncertainties among the different instruments.}. The best fitting parameters for the average emission are an absorption column density of $N_{\rm H}=(1.95\pm0.03)\times10^{20}$ cm$^{-2}$ and a photon index of $\Gamma=1.65\pm0.01$, giving an unabsorbed luminosity of $L_X=(5.57\pm0.09)\times10^{33}$\,\lum\ over the energy range 0.3--50\,keV ($\chi^2_{{\rm red}} = 1.18$ for 173 degrees of freedom).
The spectra associated with the high X-ray mode were also modelled so as to derive the SED of the pulsed emission (see Sect.\,\ref{SED}). In the fit, the column density was held fixed to the average value. We derive $\Gamma=1.66\pm0.01$ and $L_X=(6.44\pm0.01)\times10^{33}$\,\lum\ (0.3--50\,keV; $\chi^2_{{\rm red}} = 1.14$ for 199 degrees of freedom). 

\begin{table*}
\renewcommand{\arraystretch}{1.2}
\centering
\caption{Summary of the observations of PSR J1023.}              
\label{table:1}      
\begin{tabular}{l l c c c c c}          
\hline\hline                        

Telescope & Detector & Mode & Filter/Band & Start time &Net exposure & Orbit \\
 & & & &[MJD(TDB)] & [s] & \\
\hline
HST & STIS/NUV-MAMA & Fast timing spectroscopy & G230L & 59369.00660 & 2060.0 & 1\\
& &  &  & 59369.06656 & 2689.9 & 2 \\
\textit{XMM-Newton} & EPIC pn & Fast timing & 0.3--10\,keV & 59368.86453 & 58659.1 & --\\ 
\textit{XMM-Newton} & EPIC MOS1 & Full frame & 0.3--10\,keV & 59368.88195 & 58066.2 & --\\ 
\textit{XMM-Newton} & EPIC MOS2 & Full frame & 0.3--10\,keV & 59368.93910 & 53466.1 & --\\ 
NuSTAR & FPMA & Fast timing & 3--79\,keV & 59368.86938 & 22303.5 &--\\
NuSTAR & FPMB & Fast timing & 3--79\,keV & 59368.86938  & 22148.6 &--\\
\hline
HST &STIS/NUV-MAMA & Fast timing spectroscopy & G230L & 57917.63509 & 2317.0 & 1 \\
& &  &  &  57917.70131 & 2402.5 & 2 \\
& &  &  &  57917.76754 & 2058.1 & 3 \\
& &  &  &  57917.83376 & 1960.8 & 4 \\
 \textit{XMM-Newton} & EPIC pn & Fast timing & 0.3--10\,keV& 57917.60440 & 20737.9 & --\\ 
 \hline
HST & STIS/NUV-MAMA & Fast timing spectroscopy & G230L & 56774.61561 & 2348.0 & 1 \\
& &  &  &  56774.67383 & 2978.9 & 2 \\
& &  &  &   56774.74021 & 2978.9 & 3 \\
\textit{XMM-Newton} & EPIC pn & Fast timing & 0.3--10\,keV & 56818.18006 & 116712.7 & --\\ 
\hline

\end{tabular}
\end{table*}

 \section{Timing analysis}
 \label{Timing}
 
In order to perform a high-precision timing analysis, we must take the motion of the pulsar around the barycentre of the binary into account. The consequence of the orbital motion is the so-called R{\o}mer delay, which
causes a periodic anticipation or delay of the ToAs,
depending on the position of the pulsar along the orbit. For binary systems with a nearly circular orbit such as J1023 ($e < 2 \times 10^{-5}$; \citealt{Archibald}), the R{\o}mer delay equation can be simplified as  $\Delta_{R} =x \ \sin{M_{an}}$ \citep{Blandford1976}, where $x$ is the projected semi-major axis of the pulsar orbit and $M_{an}$ is the mean anomaly. The mean anomaly for a circular orbit can be written simply as $M_{an}\equiv\Omega_b$($t$-$\Tasc$), where $\Omega_b$ is the angular velocity and $\Tasc$ is the epoch of passage at the ascending node.
As reported by \cite{Jaodand} and \cite{Burtovoi}, J1023 shows large variability in the orbital phase during both the sub-luminous disc state and the radio pulsar state. This unpredictable variability leads to a large uncertainty in $\Tasc$, jeopardising the pulse search. To overcome this effect, we performed a search in the value of $\Tasc$ for each \textit{XMM-Newton} observation by varying this parameter in an interval of width $\pm200$\,s around the value extrapolated from \cite{Jaodand}. We used a step size of 0.125\,s for $\Tasc$ and folded the time series with $n$ = 16 phase bins. The final best values of $\Tasc$ were determined by fitting the peak of the $\chi^2$ distribution with a Gaussian function. 
The best $\Tasc$ and rotational period for each X-ray observation are reported in Table\,\ref{table:2}. 
To refine the measurement of the rotational period, we used the phase fitting method or the epoch folding search \citep[EFS;][]{Leahy1983, Leahy1987} with 16 phase bins when the UV signal was too weak to be detected. The EFS technique  consists of searching for periodicities in a time series by folding data over a range of trial frequencies around the candidate NS spin frequency and determining the $\chi^2$ for each folded time series.

We corrected the NuSTAR ToAs for the orbital motion using the $\Tasc$ measured from the 2021 simultaneous \textit{XMM-Newton} observation. We performed a new EFS of the NuSTAR data around the expected spin period of 1.6879874446\,ms due to the drift of the NuSTAR internal clock \citep{Madsen2015}. The pulse profile with
the largest signal-to-noise ratio corresponds to the period $P_{\rm NuSTAR} =1.687987195(21)$\,ms, which differs by $-2.5\times 10^{-10}$\,s from $P_{\rm XMM}$.

\subsection{UV timing analysis}

We expected the value of $\Tasc$ in the UV band to be very close to the one derived from the analysis of X-ray data. We performed a new search in the UV band due to the uncertainty of the NUV-MAMA detector on the absolute timing of the HST data ($\sim$1\,s; HST helpdesk private communication). We searched for the best value of $\Tasc$ in an interval of $\pm3$\,s around the best X-ray value with a step size of 0.125\,s.
The $\Tasc$ search was carried out by merging the different HST orbits from each of the three observing epochs into three corresponding chunks to increase the signal-to-noise ratio. Then we repeated the analysis for each HST orbit. We find that the pulsed signal is not detected in each observation. 

\begin{figure}
   \centering
   \resizebox{\hsize}{!}{\includegraphics{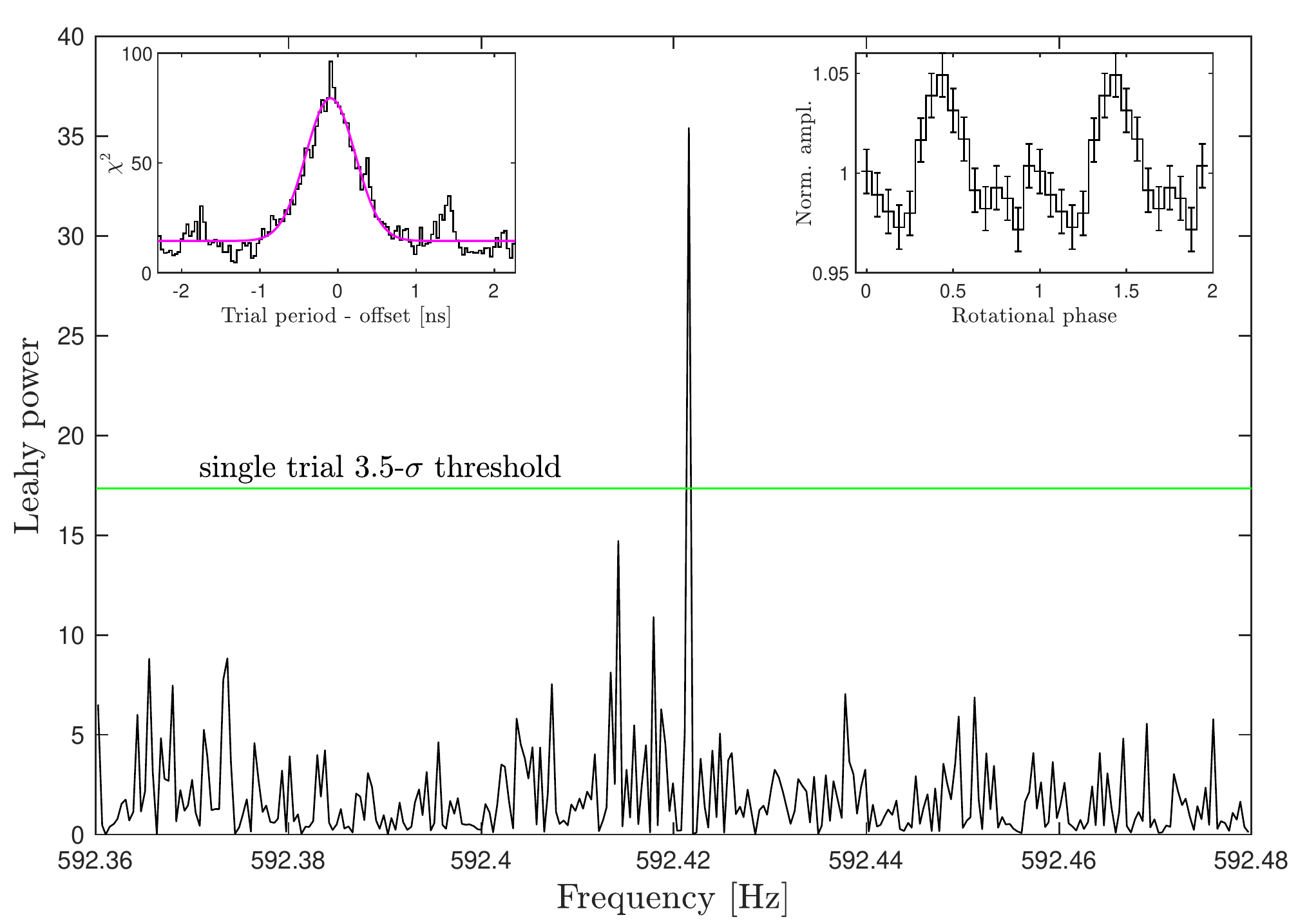}}
   \centering
      \caption{Fourier
power spectral density of the 165--310\,nm photons observed by STIS during the second orbit of the 2021 HST observations. Times of arrival are corrected using $\Tasc$\,=\,59368.091163(33)\,MJD. The horizontal green line marks the detection power threshold at 3.5$\sigma$ single trial probability. The left inset plot shows the EFS with 128 trial periods conducted by sampling each period with 16 phase bins. The right inset plot shows the folded light curves using 16 phase bins per period. The time series is folded with the spin period of 1.687987321(69)\,ms.}
         \label{PS}
   \end{figure}

We determined the $\Tasc$ value in the second HST orbit of the 2021 observations and in all merged orbits of the 2017 observations after selecting the high-mode intervals (see Fig.\,\ref{LC}). 
We computed the Fourier power spectral density for the 2021 observation after correcting the ToAs with the best $\Tasc$\,=\,59368.091163(33)\,MJD. We measure a Leahy normalised power of 35.4 at a frequency of 592.42157(20)\,Hz (Fig.\,\ref{PS}). The probability associated with random white noise fluctuations is $p = 2.0 \times 10^{-8}$, corresponding to a detection at the 5.6$\sigma$ level (single trial probability). In the insets of Fig.\,\ref{PS} we report the EFS on the left and the UV pulse profile on the right. The EFS was performed  with 128 trial periods by sampling
each period with 16 phase bins. The best spin period is 1.687987321(69)\,ms, which differs by -1.2$\times10^{-10}$\,s from $P_{XMM}$. The background-subtracted rms amplitude of the UV pulsation in the high-mode intervals is (2.13$\pm$0.36)\% (see Table\,\ref{table:3}). The background count rate was estimated by selecting photons in the channels 200--800 and 1200--1800 of the slit to have high statistics and avoid contribution from the source. We then normalised the average count rate to the total number of slit channels.

After selecting the high-mode intervals in the 2017 dataset, we merged the four orbits; we detect UV pulsations at the 3.8$\sigma$ level, considering 64 trials in the EFS. The best $\Tasc$ and rotational period are reported in Table\,\ref{table:2}. The difference between the UV and X-ray rotation period is 1.9$\times 10^{-11}$\,s. The background-subtracted rms amplitude of the UV pulsation is (1.10$\pm$0.32)\%, lower than the value measured in the 2021 observation. Similar results were obtained by \cite{Jaodand2021}; the slight differences are probably due to the different selection of the high luminosity modes. In our analysis, high modes are selected in all orbits, including the third and fourth orbits, where the flaring activity is likely to dominate. We only excluded the last part of the fourth HST orbit because it does not overlap with the X-ray observation. Flares could originate from a mechanism unrelated to low-high mode transition, such as from the outer region of the accretion disc. This would suggest that, even during the time intervals of flaring activity, J1023 does not stop the low-high mode transition; therefore, we also included the last two orbits in the analysis. We find that the highest pulse amplitude, $(1.56\pm0.59)\%$, is in the third HST orbit, despite the presence of flares (see Table\,\ref{table:3} for all UV rms pulse amplitudes). Selecting the flaring mode intervals, the double-peaked pulse profile is still visible when the time series of the third HST orbit is folded, but the significance is lower than 1.5$\sigma$, probably due to the poor statistics. As a result, we do not also rule out the presence of UV pulsations during flares.

In the 2014 observations, the signal is too weak to allow an estimate of the best $\Tasc$ value. For this reason, we corrected the ToAs for the orbital motion using the $\Tasc$ measured from X-ray observations performed 21 days later. We propagated it at the time of the UV observations (see Table\,\ref{table:2}) and then searched for the pulsed emission. Even when combining the high modes of all three orbits, the UV pulsation is not significant, but the double-peaked pulse profile is visible.

 We find a large variability in the UV rms amplitude over the time intervals of the high mode. In the two 2021 observations, both $\sim$2\,ks long, the rms amplitude of the signal changes. In the first HST orbit, at orbital phases within the range 0.62--0.74, the signal is not detected. 
 On the other hand, in the second HST orbit, in the orbital phase range 0.92--1.07, the UV signal is detected with an rms amplitude of (2.13$\pm$0.36)\%. This is the largest rms amplitude measured in the UV band. The average source and background count rates are comparable in the two orbits, and no flaring activity is detected in either observation. The orbital phase range is the only difference between the two observations. Unfortunately, the few observations in the UV band do not allow us to study a correlation between the pulse amplitude and the orbital phase. A large variability in the pulse amplitude is also observed in the optical band. The optical rms pulse amplitude during the high mode is not constant in time, ranging from values lower than 0.2\% up to a maximum value of $\sim$1.5-2\%, without a clear correlation with the orbital phase \citep{Papitto2019}.

\subsection{UV upper limits}
\label{UL}

Ultraviolet pulsations are not detected during intervals of low mode, nor during some high luminosity modes. Table\,\ref{table:3} reports the upper limits on the UV rms amplitude derived from the stack of all low-mode periods for each of the three datasets and from the high-mode intervals where the signal is not detected. Upper limits were estimated by computing the Fourier power spectral density for each observation and measuring the power of the first (P$_1$) and second (P$_2$) harmonics of the spin frequency. We then converted these two powers into rms amplitudes at the 3$\sigma$ confidence level following the procedure reported in \cite{Vaughan1994}.

To check the results, we also estimated upper limits through simulations carried out with the \texttt{Stingray} software package\footnote{\url{https://github.com/StingraySoftware/stingray}} \citep{Huppenkothen2019_1, Huppenkothen2019_2}. For each dataset, we generated 10$^4$ synthetic light curves, keeping the same integration time and mean count rate as our observations. We computed the Fourier power spectral density for each curve and measured the power of the first (P$_1$) and second (P$_2$) harmonic of the spin frequency. We then built the distribution of both P$_1$ and P$_2$ to obtain the two
values that exceed the observed powers with a high confidence level (3$\sigma$). We then converted these two powers into rms amplitudes. The results are comparable with those obtained following the procedure presented in \cite{Vaughan1994}.

The upper limits on the pulse amplitude in the low-mode intervals are 2.7\%, 2.0\%, and 1.8\% (see Table\,\ref{table:3}). Upper limits of 2.3\% and 2.4\% were set on the pulse amplitude during high-mode intervals in which UV pulses are not detected. During the flaring mode, the UV pulses are not detected down to an upper limit of 1.6\%.

\subsection{X-ray pulse amplitude variability}

Given the random evolution of both the UV and optical rms pulse amplitudes, we also investigated the variability in the X-ray band. We selected the high-mode intervals in the \textit{XMM-Newton} observations from 2021 and 2017 and divided the time series into intervals of $\sim$2.4\,ks each (see Table\,\ref{table:4}). We measured the rms pulse amplitude in all intervals to search for a possible orbital phase dependence. 

\begin{figure}
   \centering
   \resizebox{\hsize}{!}{\includegraphics{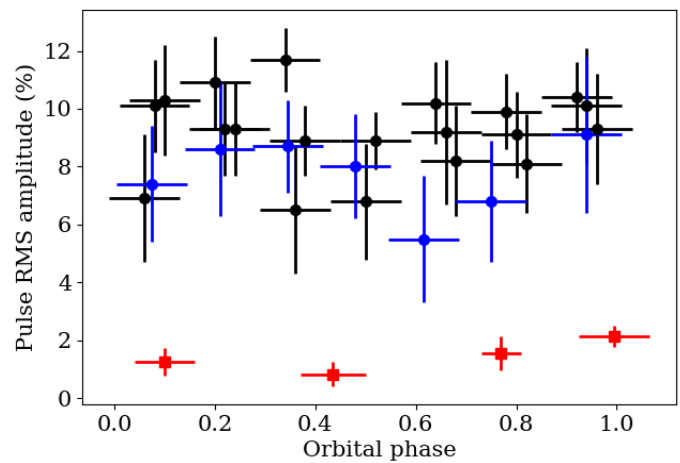}}
   \centering
      \caption{Root mean square amplitude of the pulse as a function of the orbital phase for the \textit{XMM-Newton} observations in 2017 (blue points) and 2021 (black points), and the HST observations in 2017 and 2021 (red squares).}
         \label{phase_plot}
   \end{figure}

We find a variability in the rms pulse amplitude ranging from (5.5$\pm$2.2)\% to (11.7$\pm$1.1)\%, without any correlation with the orbital phase (see Fig.\,\ref{phase_plot}). Although this study covers about four orbital phases, it should be extended to all \textit{XMM-Newton} observations to increase the statistics because of the large and random source variability.

Finally, we compared the X-ray and UV rms pulse amplitudes at almost the same orbital phase for the 2017 and 2021 observations (see Tabs.\,\ref{table:3}-\ref{table:4}). In the 2017 observations, we compared three orbital phase intervals ($\sim$0.04--0.16, 0.37--0.50, and 0.73--0.82), corresponding to the first three HST orbits. In the first two intervals, the UV pulsed signal is very faint, while in the X-ray band the rms pulse amplitude is almost constant at $\sim$7-8\%. In the third interval, dominated by flares, the UV pulse is detected with an rms pulse fraction of 1.56(59)\% and the X-ray pulsations are barely detected. 
In the 2021 observation, the X-ray rms amplitudes in the phase intervals 0.62--0.74 and 0.92--1.07 are compatible with each other within the errors, while the UV rms pulse amplitude in the first observation is lower than in the second. Notwithstanding the rather low statistics, we notice that pulsed amplitude variations may not occur simultaneously at different frequencies.

\begin{table*}
\renewcommand{\arraystretch}{1.2}
\centering
\caption{Summary of the best times of the ascending node passage obtained from folding the \textit{XMM-Newton} and HST time series by varying the $\Tasc$. In the last column we report the best value of the rotational period obtained by the phase fitting technique or by fitting the EFS periodograms with a Gaussian.}              
\label{table:2}      
\begin{tabular}{c c c c}          
\hline\hline                        
Obs. ID & Obs. start & Best $\Tasc$ [MJD] & Best $P_{\rm s}$ [ms]\\
\hline
\multicolumn{4}{c}{\textit{XMM-Newton}} \\
\hline
0864010101 & Jun 3, 2021 & 59368.0910979(27) & 1.68798744494(51)\\
0803620501 & Jun 13, 2017 & 57917.431286(12) & 1.6879874534(51)\\
0742610101 & Jun 10, 2014 & 56818.1948362(24) & 1.68798744496(24)\\
\hline
\multicolumn{4}{c}{\textit{Hubble Space Telescope}}\\
\hline
16061 & Jun 4, 2021 & 59368.091163(33) & 1.687987321(69)\\
14934 & Jun 13, 2017 & 57917.431293(79) & 1.687987472(44)\\  
13630 & Apr 27, 2014 & 56774.6136468(24)$^{*}$& -- \\
 \hline

\end{tabular}\\
$^{*}$ Value propagated by the closest X-ray ephemeris (Obs. ID 0742610101).\\
\end{table*}

\begin{table}
\renewcommand{\arraystretch}{1.2}
\centering
\caption{Properties of the UV pulses. The total rms amplitude is estimated from the rms amplitude of the first and second harmonic ($A_1$ and $A_2$) as $A=(A_1^2+A_2^2)^{1/2}$. The uncertainties on the rms amplitudes are at a confidence level of 1$\sigma$, while the upper limits on the rms amplitudes are at a 3$\sigma$ confidence level.}              
\label{table:3}      
\begin{tabular}{l c c c c}          
\hline\hline                        
Mode & Year & Orbit & Orbital phase& Amplitude (\%) \\
\hline
High & 2021 & 1&0.62--0.74 & <2.3  \\
High & 2021 & 2& 0.92--1.07 &  2.13(36)\\
Low & 2021 & 1-2&0.64--1.12 & <2.7  \\
High& 2021 & 1-2&0.62--1.07 & 1.41(23)\\
\hline
High& 2017 & 1&0.04--0.16 & 1.25(47)$^*$  \\
High& 2017 & 2&0.37--0.50 & <2.4\\
High& 2017 & 3&0.73--0.82 & 1.56(59) \\
High& 2017 & 1-2-3-4$^{**}$ & 0.04--1.08 & 1.06(37) \\
Low& 2017 & 1-2-3&0.03--0.82 & <2.0\\
Flaring& 2017 & 3&0.76--0.81 &  <1.6 \\
\hline
High& 2014 & 1-2-3& 0.01--0.81 &  0.68(19)$^*$ \\
Low& 2014 & 1-2-3& 0.01--0.76 & <1.8  \\
\hline

\end{tabular}
$^*$ The significance of the signal is lower than 3.5$\sigma,$ but the morphology of the pulse profile is similar to the expected one.\\
$^{**}$ In the fourth orbit we consider only 50\% of the observation, where there is the overlap with the X-ray band. 
\end{table}

\begin{table}
\renewcommand{\arraystretch}{1.2}
\centering
\caption{Properties of the X-ray pulses in the 2021 and 2017 observations after selecting the high luminosity mode intervals and dividing the observation into intervals of 2.4\,ks. The total rms amplitude is estimated from the rms amplitude of the first and second harmonics ($A_1$ and $A_2$) as  $A=(A_1^2+A_2^2)^{1/2}$. The uncertainties are at a confidence level of 1$\sigma$.}              
\label{table:4}      
\begin{tabular}{l c c}          
\hline\hline                        
Start time [MJD] & Orbital phase& Amplitude (\%) \\
\hline
59368.920088       & 0.18--0.33 &   -- $^{*}$\\
59368.947866       & 0.33--0.47 &   -- $^{*}$\\
59368.975644       & 0.47--0.61 &   4.4 $\pm$ 2.2$^{*}$\\
59369.003421       & 0.61--0.75 &   8.2 $\pm$ 1.9\\
59369.031199       & 0.75--0.89 &   8.1 $\pm$ 1.7\\
59369.058977       & 0.89--0.03 &   9.3 $\pm$ 1.9\\
59369.086755       & 0.03--0.17 &   10.3 $\pm$ 1.9\\
59369.114533       & 0.17--0.31 &   9.3 $\pm$ 1.6\\
59369.142310       & 0.31--0.45 &   8.9 $\pm$ 1.2\\
59369.170088       & 0.45--0.59 &   8.9 $\pm$ 1.0\\
59369.197866       & 0.59--0.73 &   9.2 $\pm$ 2.5\\
59369.225644       & 0.73--0.87 &   9.1 $\pm$ 1.5\\
59369.253421       & 0.87--0.01 &   10.1 $\pm$ 2.0\\
59369.281199       & 0.01--0.15 &   10.1 $\pm$ 1.6\\
59369.308977       & 0.15--0.29 &   9.3 $\pm$ 1.6\\
59369.336755       & 0.29--0.43 &   6.5 $\pm$ 2.2\\
59369.364533       & 0.43--0.57 &   6.8 $\pm$ 2.0\\
59369.392310       & 0.57--0.71 &   10.2 $\pm$ 1.4\\
59369.420088       & 0.71--0.85 &   9.9 $\pm$ 1.3\\
59369.447866       & 0.85--0.99 &   10.4 $\pm$ 1.2\\
59369.475644       & 0.99--0.13 &   6.9 $\pm$ 2.2\\
59369.503421       & 0.13--0.27 &   10.9 $\pm$ 1.6\\
59369.531199       & 0.27--0.41 &   11.7 $\pm$ 1.1\\
\hline
57917.604397          & 0.87--0.01 & 9.1 $\pm$ 2.7\\
57917.631064          & 0.01--0.14 & 7.4 $\pm$ 2.0\\
57917.657730          & 0.14--0.28 & 8.6 $\pm$ 2.3\\
57917.684397          & 0.28--0.41 & 8.7 $\pm$ 1.6\\
57917.711064          & 0.41--0.55 & 8.0 $\pm$ 1.8\\
57917.737730          & 0.55--0.68 & 5.5 $\pm$ 2.2\\
57917.764397          & 0.68--0.82 & 6.8 $\pm$ 2.1\\
57917.791064          & 0.82--0.95 & 4.7 $\pm$ 1.9$^{*}$\\
57917.817730          & 0.95--0.09 & -- $^{*}$\\
\hline

\end{tabular}

$^*$ The pulsed signal is absent or barely significant due to the presence of flares.
\end{table}

\section{Spectral energy distribution}
\label{SED}
Figure\,\ref{SED} shows the SED of the total and pulsed emissions in the UV and X-ray bands during the observations in June 2021. The HST/near-UV spectrum is corrected for the interstellar extinction, considering the absorption column $N_H= 1.9 \times 10^{20}$\,cm$^{-2}$ measured by X-ray spectral fitting (see Sect.\,\ref{X-ray}) and the empirical relation $A_V = N_H/(2.87 \pm 0.12) \times 10^{21}$\,cm$^{-2}$ \citep{Foight}. The resulting colour excess is $E(B-V)=0.022$. The total UV fluxes were calculated by integrating the spectrum within the 165--275\,nm and 275--310\,nm intervals. These values were multiplied by the rms UV pulse amplitudes 
to obtain the UV pulsed fluxes. In the X-ray band we evaluated the values for the background-subtracted pulse rms amplitudes over the energy ranges 0.3--2, 2--5, 3--10, and 5--10\,keV and the 3$\sigma$ upper limit over the range 10--50\,keV. We estimated the unabsorbed X-ray fluxes for the pulsed plus un-pulsed emission in the high mode over the same energy ranges and multiplied them by the corresponding values of the pulse rms amplitude to obtain integrated X-ray pulsed fluxes. To convert the pulsed X-ray fluxes (and upper limits) into $\nu$\,$F_{\nu}$ units, we multiplied the fluxes by the ratio between the mid-point energy of the interval and the width of the energy interval. We then fitted the pulsed SED with a power-law model, obtaining a functional dependence of the form $\nu$\,$F_{\nu}= \nu^{0.44\pm 0.04}$ (uncertainty at the 90\% confidence level). This power-law relation is similar to that found by \cite{Papitto2019} ($\nu$\,$F_{\nu} \sim \nu^{0.3}$), which connected the 320--900\,nm optical  and the 0.3--45 keV X-ray bands. The slight discrepancy between the two results is due to the facts that we used two different values of $N_H$ and that the pulsed amplitude is variable over time.

\begin{figure}
   \centering
   \resizebox{\hsize}{!}{\includegraphics{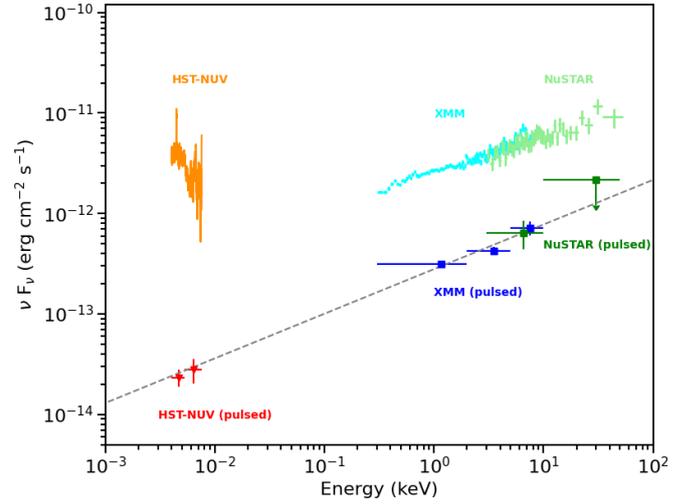}}
   \centering
      \caption{Unabsorbed SED from the near-UV to X-rays obtained using data collected simultaneously with HST, \textit{XMM-Newton}, and NuSTAR during the 2021 simultaneous multi-wavelength campaign. The HST spectrum is plotted with an orange line and ranges from 165 to 310\,nm. The average pulsed near-UV flux is plotted with red triangles over the 165--275\,nm and 275--310\,nm intervals. The total (pulsed plus un-pulsed) X-ray fluxes measured using \textit{XMM-Newton} and NuSTAR are plotted using light blue and green points, respectively. The pulsed X-ray fluxes measured using \textit{XMM-Newton} are computed over the 0.3--2, 2--5, and 5--10\,keV energy bands and are plotted using blue squares. The pulsed X-ray fluxes measured using NuSTAR are computed over the 3--10 and 10--50\,keV energy bands and are plotted using green squares. The upper limit in the 10--50\,keV interval is marked by an arrow. The dashed grey line indicates the best fitting power-law model $\nu$\,$F_{\nu} \sim \nu^{0.4}$ that matches the pulsed near-UV and X-ray fluxes.}
         \label{SED_plot}
   \end{figure}

\section{Discussion and conclusions}
\label{Discussion}
In this work we have presented the results of HST, \textit{XMM-Newton}, and NuSTAR observations of the tMSP J1023 carried out in 2014, 2017, and 2021 during its sub-luminous disc state. We have confirmed that the UV light curve of J1023 shows three luminosity modes (low, high, and flaring), as presented in \cite{Jaodand2021}. The transition between the UV high and low luminosity modes occurs very rapidly and simultaneously with that in the X-ray band. The average UV flux of the source is not constant between two different HST orbits. Similar variability was also observed in the optical band \citep{Papitto2018, CotiZelati2018}. For this reason, in the absence of strictly simultaneous X-ray observations, it is difficult to select the intervals of low, high, and flaring modes properly. 

We performed UV and X-ray phase-resolved timing analysis, studying the rms amplitude variability along the orbit through $\sim$2.4\,ks long time intervals. 
Coherent UV pulsations were detected only in the high luminosity mode, with a maximum and minimum rms amplitude of (2.13$\pm$0.36)\% and (0.68$\pm$0.19)\%, respectively.
Similar to optical and X-ray bands, UV pulses were not detected in the low mode. The double-peaked pulse profile typical of J1023 was visible when the UV data that correspond to the flaring mode of the third 2017 HST orbit were folded; however, it was not significant. It is possible that the UV pulsed signal could also be present during flares with an amplitude lower than during the high mode, as observed the optical band \citep{Papitto2019}. The poor statistics did not allow us to confirm the UV pulsed signal during the flaring activity.
In the X-ray band during the high luminosity mode, the rms amplitude ranges from $\sim5.5\%$ to $\sim12\%$, without any evident correlations with the orbital phase. In different previous studies (\citealt{Archibald2015, Jaodand, Papitto2019}), X-ray
pulsations were observed with an average
rms amplitude of $\sim$8\%, in agreement with our mean value of (7.41$\pm$0.49)\%. To definitively rule out the orbital phase dependence, the study should be extended to all \textit{XMM-Newton} observations. Unfortunately, in the UV band the statistics are too low to search for a possible correlation with the orbital phase.

J1023 and the accreting millisecond X-ray pulsar SAX\,J1808.4--3658 are the only two MSPs observed to emit optical and UV pulses \citep{Ambrosino, Papitto2019, Ambrosino2021}. These recent discoveries challenged our understanding of the physical processes that generate both X-rays and UV/optical pulses in binaries. \cite{Papitto2019} suggested that the optical and X-ray pulsations of J1023 are produced just outside
the light cylinder, where the accretion disc meets the striped pulsar wind. Optical and X-ray pulses are produced almost simultaneously (time lag of $\sim 200\,\mu$s; \citealt{Papitto2019}, Illiano et al. in prep.) by synchrotron emission in the intra-binary shock \citep{Veledina, Campana2019}. In fact, the optical and X-ray pulses had a similar pulse profile, and the pulsed flux spectral distribution from the optical band to 45\,keV was compatible with a power-law relation $\nu F_{\nu} \sim \nu^{0.3}$. In this work we found a similar result by connecting the pulsed UV flux with that in the X-ray band up to $\sim$50\,keV. Figure \ref{SED_plot} shows the pulsed flux spectral distribution of the 2021 simultaneous observations. 
This result in the UV band supports the hypothesis of a unique non-thermal emission process for X-ray, UV, and optical pulses.

It is not yet clear why there is a strong variability in the pulsed fraction that is not correlated with the orbital phase and between different wavelengths. A decrease in the UV pulsed fraction does not always correspond to a decrease in the X-ray pulsed fraction, and vice versa. 
Variations in the pulse amplitude are probably related to small changes in
the mass accretion rate leading to a random expansion and contraction of
the intra-binary shock region. In the low mode, this region may expand in a way that makes the pulsation disappear completely. In the high mode, variations in the mass accretion rate are smaller than in the low mode, but they can still generate random oscillations in the pulse amplitude. A decrease in the accretion rate during the high mode can determine the expansion of
the shock region. In this condition, part of the energy carried by the NS may
be lost instead of being absorbed by the shock region and re-emitted through the
synchrotron mechanism. At the same time, an increase in the distance of the intra-binary shock region from the NS, caused by a decrease in the mass accretion
rate, can lead to the loss of signal coherence. The electrons accelerated in the shock radiate their energy via synchrotron emission with a timescale $t_{\rm sync} \propto \epsilon^{-1/2} B_s^{-3/2}$, where $\epsilon$ is the photon energy and $B_s$ is the post-shock magnetic field. The synchrotron timescale increases if the shock is located at a larger distance because
the strength of the post-shock magnetic field decreases linearly
with distance. When $t_{sync}$ becomes
comparable to half the NS spin period, we expect to observe a total or partial loss of the signal coherence. This occurs for UV photons ($\sim$4 eV) when the magnetic field at the shock is as low as $1 \times 10^5$\,G. In the X-ray band, the $t_{\rm sync}$ is shorter than in the UV and optical bands, so we expect the loss of signal coherence to be less pronounced. If changes in the size and location of the shock are at the origin of the variations in the pulse amplitude, we expect to observe a correlation between neighbouring bands, such as optical and UV. In the future, simultaneous observations in the far-UV, near-UV, and optical bands would allow us to probe this hypothesis by studying the correlation between variations in the pulse amplitude in different bands.

\begin{acknowledgements}
We thank the referee for useful comments. Part of this paper is based on observations with the NASA/ESA Hubble Space Telescope, obtained at the Space Telescope Science Institute, which is operated by AURA, Inc., under NASA contract NAS5-26555. 
Some of the scientiﬁc results reported in this study are based on observations obtained with; \emph{XMM-Newton}, which is an European Space Agency (ESA) science mission with instruments and contributions directly funded by ESA member states and NASA; and the NuSTAR mission, which is a project led by the California Institute of Technology, managed
by the Jet Propulsion Laboratory and funded by NASA. The \emph{XMM-Newton} SAS is developed and maintained by the Science Operations Centre at the European Space Astronomy Centre. The NuSTAR Data Analysis Software (NuSTARDAS) is jointly developed by the ASI Science Data Center (ASDC, Italy) and the California Institute of Technology (Caltech, USA). This research has made use of softwares and tools provided by the High Energy Astrophysics Science Archive Research Center (HEASARC) Online Service.
    AMZ is supported by PRIN-MIUR 2017 UnIAM (Unifying Isolated and Accreting Magnetars, PI S. Mereghetti). 
    FCZ is supported by a Juan de la Cierva fellowship (IJC2019-042002-I).
\end{acknowledgements}

%
%

\bibliography{biblio}

\end{document}